%% file: main.tex
\documentclass{book}

\usepackage{makeidx}
\usepackage{amsthm,amsmath,amssymb,amsfonts}
\usepackage{setspace,graphicx}
\usepackage{color}
\usepackage{xcolor}
\usepackage{Generic}  
\usepackage[sort,longnamesfirst]{natbib}
\usepackage{url}
\usepackage{multirow}
\usepackage{fancyvrb}
\usepackage{booktabs}
\usepackage[shortlabels]{enumitem}
\usepackage{cancel}

\input{preamble.tex}

\numberwithin{equation}{section}
\theoremstyle{plain}

\begin{document}
\doublespacing
%


\chapter[Normalizing Functions]{Algorithms for Models with Intractable Normalizing Functions} \label{sec: label}

\begin{center}
\begin{large}
{\em Murali Haran, Bokgyeong Kang, Jaewoo Park
}
\end{large}
\end{center}

In this paper we discuss a well known computing problem---inference for models with intractable normalizing functions. Models with intractable normalizing functions arise in a wide variety of areas, for instance network models, models for spatial data on lattices, spatial point processes, flexible models for count data and gene expression, and models for  permutations \citep{lu2014MallowsModel}; for more examples see \citet{matsubara2022robust} and \citet{parkharan:intractable}. 
Simulating from these models for fixed parameter values is well studied, starting with work dating back seventy years to the origin of the Metropolis algorithm. On the other hand some of the most practical and theoretically justified algorithms for inference, particularly Bayesian inference, have only been developed within the past two decades. The most computationally efficient algorithms often do not have well developed theory and few if any approaches exist for assessing the quality of approximations based on them. For many problems even the best algorithms can be  computationally infeasible. Hence, this is  an exciting area of research with many open problems. We explain several key algorithms, providing connections and touching upon practical advantages and disadvantages of each, with some discussion of theoretical properties where they impact practice. We discuss an approach for assessing the accuracy of approximations produced by these algorithms; this diagnostic is particularly valuable for algorithm tuning. 

While our focus is largely on models with intractable normalizing functions, we also discuss algorithms that are more broadly applicable to models where the entire likelihood function is intractable; these methods are of course also applicable to intractable normalizing function problems. Intractable likelihood function problems are growing in importance  as statistical models become increasingly sophisticated and data sets become larger and more complex. Inference for intractable likelihood models, also known as likelihood-free inference, therefore represents one of the most important computational challenges of modern statistics. The goals of this manuscript are to (i) provide an accessible introduction to the intractable normalizing function (INF) and intractable likelihood (IL) problems, as well as key ideas underpinning several algorithms used to solve them; (ii) describe an approach for assessing the sample quality of asymptotically inexact algorithms for INF problems; (iii) provide  practical recommendations for solving INF problems based on a study of some challenging examples; (iv) suggest avenues for future research. The remainder of this paper is organized as follows. We provide some historical and technical background for INF and IL problems in Section \ref{sec:Introduction}. In Section \ref{sec:Algs} we provide a taxonomy for algorithms, along with explanations for several important algorithms, then explain an approach for assessing sample quality for both asymptotically exact and inexact algorithms in Section \ref{sec:Assess}. We conclude with the application of several algorithms to challenging examples in Section \ref{sec:Applications}, providing insights about the algorithms whenever possible, and then provide a summary and discussion of potential areas for future research in Section \ref{sec:Summary}. 

\section{Background}\label{sec:Introduction}
The landmark \cite{metropolis1953} paper that introduced the Metropolis algorithm and hence the beginnings of Markov chain Monte Carlo (MCMC), describes an algorithm to simulate a spin system in particle physics according to the Ising model \citep{ising1925beitrag,lenz1920beitrag}, which has an   intractable normalizing function. Because the problem considered in the paper is simulating from the model for fixed parameters, the normalizing function (really a constant in this case) cancels out in the acceptance ratio of the Metropolis-Hastings algorithm. 
Hence, from the earliest days of MCMC, simulating realizations from a model with intractable normalizing functions has been considered a solved problem.
However, inference is an entirely different matter as the normalizing function does not cancel out. We explain below both the ease of simulation from the model and the challenge for inference. 

Consider a random variable $x\in \mathcal{X}$, assumed to be a realization from the probability model $f(x\mid\theta)$ with parameter  $\theta\in\Theta$. Bayesian inference for $\theta$ is based on the posterior $\pi(\theta\mid x)\propto f(x\mid\theta)p(\theta)$ where $p(\theta)$ is a prior density on $\theta$. It is common to have $f(x\mid\theta) = h(x\mid\theta)/c(\theta)$ where $c(\theta)$ is an intractable normalizing function of $\theta$. For instance, Ising, Potts, and exponential random graph models (ERGMs) can  be expressed as exponential family models of the form $f(x\mid \theta) = \exp(\theta S(x))/c(\theta),$ where $\theta$ is a $p$-dimensional vector of parameters and $S(x)$ is the vector of jointly sufficient statistics for $\theta$. The normalizing function $c(\theta) = \sum_{x\in \chi} \exp(\theta s(x))$ 
is intractable for any realistic problem as it involves summing over all possible  $x\in \mathcal{X}$. For example for the Ising model which involves 2 potential spins $\{-1,1\}$ for each of $n$ particles, the number of configurations is $2^n$, a large number even for relatively small $n$; for $n=300$ it is more than $2\times 10^{90}$ configurations. 

The posterior distribution for the kinds of models described above is $\pi(\theta\mid x)\propto h(x\mid\theta)p(\theta)/c(\theta)$. This reveals the source of the challenge with constructing a Metropolis-Hastings sampler for $\pi(\theta\mid x)$: if the current state of the Markov chain is $\theta^{t}$ then the proposed state $\theta^* \sim q(\theta^{t}, \cdot)$, for some proposal $q(\cdot,\cdot)$, is accepted with probability $$\min\left(1, \frac{p(\theta^*)h(\theta^* \mid x)/c(\theta^*) }{p(\theta^t)h(\theta^{t} \mid x)/c(\theta^{t})} \frac{q(\theta^*, \theta^{t})}{q(\theta^{t}, \theta^*)} \right),$$
where the intractable normalizing function does not cancel out. 

In contrast, simulating from the probability model is itself quite straightforward in principle. That is, for a fixed $\theta$, a Metropolis-Hastings update for sampling from the model $f(x\mid\theta)$ is as follows: if the current state of the Markov chain is $x^{t}$ then the proposed state $x^* \sim q(x^{t}, \cdot)$ for some proposal $q(\cdot,\cdot)$ is accepted with probability $$\min\left(1, \frac{h(x^* \mid\theta)/c(\theta)  }{h(x^{t}\mid\theta)/c(\theta)} \frac{q(x^*, x^{t})}{q(x^{t}, x^*)} \right) = \min\left(1, \frac{h(x^* \mid\theta)}{h(x^{t}\mid\theta)} \frac{q(x^*, x^{t})}{q(x^{t}, x^*)} \right),$$
which does not depend on the intractable normalizing function $c(\theta)$. This simple observation allows for easy simulation from the probability model, as first shown in \cite{metropolis1953}. As we will discuss below, simulation from $f(y\mid\theta)$ is key to most algorithms for simulating from $\pi(\theta \mid x)$. It is important to note that while the normalizing function poses no challenge to simulation, slow mixing is a common issue, 
 inspiring many innovations, notably the Swendsen-Wang algorithm \citep{swendsen1987nonuniversal} and variants. 


The earliest approaches, mostly focused on approximating maximum likelihood estimators, were based on pseudolikelihood approximations (Besag, 1975, Lindsay, 1988). These likelihood approximations, obtained by taking the product of the full conditional distributions of all variables, are simple and computationally expedient. However, they are limited to certain classes of spatial and network models. They do not apply, for instance, to the Conway--Maxwell--Poisson \citep{conway1961queueing} model or Mallows model \citep{mallows1957non,lu2014MallowsModel}. 
In cases where one can apply it, pseudolikelihood is often a poor approximation. For example, it can work poorly when the dependence is moderately strong for the Potts model or the autologistic model \citep[cf.][]{okabayashi2011extending,hughes2011autologistic}. 
In contrast, composite likelihoods, a general scheme for approximation derived by multiplying a collection of component likelihoods \citep[cf.][]{varin2011overview}, are much more flexible by allowing various kinds of marginal and conditional component likelihoods, and can be computationally expedient for many situations \citep[cf.][for an applications to the Potts model]{okabayashi2011extending}. Markov chain Monte Carlo maximum likelihood (MCMCML) is an elegant, theoretically justified approach to approximating MLEs \citep{geye:thom:1992} using importance sampling to approximate the likelihood function as well as the  curvature of the log likelihood. The challenges with MCMCLE are largely related to finding a good importance function that ensures that Monte Carlo errors   do not balloon \citep[see][for some strategies for ERGMs]{hummel2012improving}. 

Our focus in this chapter is on  algorithms for Bayesian inference. We do not discuss maximum likelihood estimation for intractable normalizing function models, nor do we discuss the large number of algorithms focused entirely on approximating normalizing functions or constants themselves \citep[see][]{gelman1998simulating,geyer2011importance}. 

\section{Algorithms}\label{sec:Algs}


Following the categories in \cite{parkharan:intractable} we can broadly classify  algorithms for Bayesian inference with intractable normalizing function models into the following overlapping categories: (i) auxiliary variable (AV) methods, and (ii)  likelihood function approximation (LFA) methods. AV methods typically avoid the evaluation of normalizing functions by introducing a well chosen auxiliary variable into the sampling algorithm. 
LFA methods construct an approximation to the full likelihood function and use the approximation in place of the true likelihood when evaluating the posterior distribution. These approximations can vary widely, from composite likelihood or pseudolikelihood approximations dating back to the 1970s to the use of Gaussian processes or a variety of new machine learning approaches to approximate the likelihood based on samples from $f(y\mid\theta)$ at various $\theta$ values. Likelihood function approximations include many methods that approximate the normalizing function. The categories clearly overlap heavily since most LFA algorithms require drawing samples from the data model $f(y\mid\theta)$, and hence may also be considered AV algorithms. 
In spite of the overlap, we find the above categorization to be helpful as a way to distinguish the fairly distinct thought processes behind constructing different algorithms for intractable normalizing function problems. 
The above algorithms may also be categorized as asymptotically exact or inexact: for asymptotically exact algorithms the asymptotic distribution of the stochastic process produced by the algorithm---often but not always a Markov chain---is exactly equal to $\pi(\theta \mid y)$; inexact algorithms do not have this property. 

We provide an overview of algorithms for Bayesian inference for models with intractable normalizing functions. We sprinkle in discussions of algorithms for models where the  entirety of the likelihood is intractable because, of course, algorithms for IL also apply to INF problems. 
Note that there is a vast and fast growing literature on IL methods, often referred to as ``simulation-based inference'', that spans multiple disciplines;
see \cite{cranmer2020frontier} for an authoritative review. 
As will become apparent, algorithms for sampling from $f(y\mid\theta)$ end up being  crucial to almost all algorithms developed for  INF or IL problems. 

\subsection{Auxiliary Variable  Algorithms}\label{subsec:AuxVar}

Here we describe algorithms where an additional simulation is used in various ways to cancel out normalizing function evaluations in the Metropolis-Hastings acceptance ratio. 
Both the exchange algorithm and double Metropolis-Hastings apply to INF problems. We also describe briefly the MCMC version of the Approximate Bayesian Computation approach. This algorithm is widely applicable to IL problems, entirely eliminating the need for evaluating the likelihood function. 

\subsubsection{Exchange Algorithm}
The exchange algorithm \citep{moll:pett:bert:reev:2006,Murray2007Advances-in-Mar} relies on an exact sample from the probability model $f(\cdot \mid\theta)$ to produce a Markov chain with stationary distribution $\pi(\theta\mid x)$. Let the $t$th state of the Markov chain with augmented state space be $(\theta^t, x^t)\in \Theta \times \mathcal{X}$. In the exchange algorithm, each time a parameter value $\theta^*$ is proposed from $q(\theta^t,\cdot)$, an auxiliary sample from the probability model at that parameter value $\theta^*$ is drawn, $x^* \sim f(\cdot\mid\theta^*)$. The joint proposal $(\theta^*, x^*)$ is then accepted or rejected together to construct a Markov chain on the augmented state space.  The Metropolis-Hastings acceptance ratio for the proposal $(\theta^*, x^*)$ is 
\begin{equation}
    \alpha(\theta^t, \theta^*)= \min\left\{ 1,
    \frac{p(\theta^*)h(x\mid\theta^*)/{\cancel{c(\theta^*)}}} {p(\theta^t) h(x\mid\theta^t)/\bcancel{c(\theta^t)}}\:\:
    \frac{q(\theta^*, \theta^t) h(x^*\mid\theta^t)/\bcancel{c(\theta^t)}}
        {q(\theta^t, \theta^*)  h(x^* \mid\theta^*)/{\cancel{c(\theta^*)}}}
\right\},
\end{equation}
which does not contain normalizing function evaluations.  The distribution of the marginal chain, that is, just the $\theta$ component of the resulting Markov chain, has stationary distribution $\pi(\theta\mid x)$. This is a very elegant approach to constructing an asymptotically exact sampler for the target posterior. An important requirement for this algorithm is that we have an exact draw $x^* \sim f(\cdot\mid\theta^*)$, that is, using a draw from a Markov chain with stationary distribution $f(x\mid\theta^*)$ does not suffice. For most models of interest, it is either impossible to construct an exact sampler for $f(x\mid\theta^*)$ or, in cases where it is possible, for example using perfect sampling techniques \citep{Prop:Wils:1996}, it is often computationally too expensive to be of value. In fact, in all the examples we provide in Section \ref{sec:Applications} perfect sampling is not a viable option.

\subsubsection{Double Metropolis-Hastings and Variants}

The Double Metropolis-Hastings algorithm (DMH) \citep{Liang2010} simply takes the exchange algorithm and replaces the exact draw from $f(\cdot\mid\theta^*)$ with a draw from a Markov chain with $f(\cdot\mid\theta^*)$ as its stationary distribution. The algorithm gets its name from the fact that at each iteration of the Markov chain (``outer chain") the algorithm requires another Markov chain (``inner chain") to provide the auxiliary draw. This is obviously  expensive, especially if the inner chain is long, but  removing the requirement of having an exact draw makes it far more flexible and efficient than the exchange algorithm. There are no theoretical guarantees regarding the quality of DHM samples since the theory for DMH requires both inner and outer chain lengths get large simultaneously, which is impractical. Thus, its efficiency comes at a considerable cost, namely that DMH is asymptotically inexact so standard approaches for assessing convergence \citep{fleg:hara:jone:2008} do not apply. 

As we discuss in Section \ref{sec:Applications}, DMH is both efficient and relatively easy to construct. Important ingredients for the construction of DMH include coming up with a good proposal for $\theta$ ($q(\theta,\cdot$)), and determining the length of the inner and outer chains. Improving proposals and determining a suitable outer chain length are standard issues in constructing MCMC algorithms but the inner chain length problem is specific to DMH. All three of these issues require the ability to assess the quality of samples produced by DMH; given the fact that the algorithm is asymptotically inexact usual MCMC diagnostics are not useful. This is an issue we will address more broadly in Section \ref{sec:Assess} and we discuss the implementation of DMH in challenging  examples in Section \ref{sec:Applications}. 

\subsubsection{Approximate Bayesian Computation}
 Approximate Bayesian Computation (ABC) is a very widely used class of algorithms for Bayesian inference in the presence of intractable likelihood functions \citep[see][for a glimpse of the extensive literature in this area]{sisson2018handbook}. Here we describe ABC-MCMC \citep{marjoram2003markov}, a subset of ABC algorithms. In ABC-MCMC, for current state $\theta^t$ when $\theta^{*}$ is proposed the Metropolis-Hastings acceptance ratio is evaluated only if an auxiliary sample  $x^{*} \sim f(\cdot\mid\theta^{*})$ is close to the data according to some user-specified distance $d(x,x^{*})$ and threshold $\epsilon>0$; of course the distance could be defined on a statistics $S(x^*)$, which is a fruitful area to explore the use of dimension reduction methods. If $d(x,x^{*})<\epsilon$, $\theta^{*}$ is accepted with probability 
\begin{equation}
    \alpha(\theta^t, \theta^{*})= \min\left\{ 1, 
    \frac{p(\theta^{*})} {p(\theta^t)}\:\:
    \frac{q(\theta^{*}, \theta^t)}
        {q(\theta^t, \theta^{*})}
\right\},
\end{equation}
For discrete data ($\mathcal{X}$ is a discrete space) when $\epsilon=0$, \cite{marjoram2003markov} use a simple detailed balance argument to prove that the resulting Markov chain converges to stationary distribution $\pi(\theta\mid x)$. Of course, this is an impractical requirement so the Markov chain with $\epsilon>0$ is in reality asymptotically inexact. Because of the growing number of problems where the evaluation of $f(x\mid\theta)$ is impossible even while auxiliary simulation from $f(\cdot\mid\theta)$ is relatively straightforward, ABC-MCMC and other ``likelihood-free'' algorithms have become very popular. ABC-MCMC is of course applicable to INF problems but its value in comparison to more specialized INF algorithms like DMH is still an open question. 


\subsection{Likelihood Approximation Algorithms}\label{subsec:LApprox}

Some of the earliest approaches for inference with intractable normalizing function likelihoods involve replacing the entire likelihood function by an approximation that completely avoids computing the normalizing function. This idea remains potentially very useful and opens avenues for research, particularly the use of new machine learning methods for approximation. In the interest of brevity, and because likelihood approximations are major topics in their own right, we provide only brief notes on some of the key ideas in this class of algorithms. 

\subsubsection*{Composite Likelihood}

In composite likelihood \citep{lindsay1988composite}, the likelihood function $\mathcal{L}(\theta\mid x)$ is replaced by an approximation $\hat{\mathcal{L}}_C(\theta\mid x)$ obtained by taking the product of a collection of component likelihoods. A maximum composite likelihood estimate (MCLE) is obtained by maximizing $\hat{\mathcal{L}}_C(\theta\mid x)$ with respect to $\theta$. MCLEs are a good approximation to the MLE in certain settings \citep[cf.][]{varin2011overview}. Composite likelihood methods are also easy to implement and computationally expedient. In fact, they are the only class of algorithms discussed in this paper that do not require auxiliary simulations. Pseudolikelihood approaches, where the likelihood is approximated by a product of full conditional distributions of the random variables in the model, are somewhat limited in their flexibility but composite likelihood methods can be adapted well to particular problems. A nice example is provided in  \cite{okabayashi2011extending} for the Potts model for discrete spatial lattice data. Here,  $\hat{\mathcal{L}}_C(\theta\mid x)$ is the product of the joint distributions of multiple pixels given the rest of the lattice. They find that the resulting MCLE is less statistically efficient but much faster and easier to compute than the MLE; the MLE can only be approximated using a combination of MCMC and importance sampling \citep{geyer1992constrained}. For Bayesian inference with composite likelihood, $\pi(\theta\mid x)$ is replaced with $\hat{\pi}(\theta\mid x)\propto \hat{\mathcal{L}}_C(\theta\mid x)p(\theta)$, as long as $\hat{\pi}(\theta\mid x)$ is proper. Composite likelihood for Bayes is a somewhat under-explored approach, though there are some  studies for example for spatial extreme models \citep{ribatet2009bayesian} and for calibrating complex computer models \citep{chang2015composite}. 

\subsubsection*{Intractable Normalizing Function Approximations}
There is a large literature on approximating the normalizing function using importance sampling, bridge sampling and path sampling methods  \citep{gelman1998simulating}. These algorithms can all be used to approximate the Metropolis-Hastings acceptance ratio. For instance,   
\citet{atch:lart:robe:2008} construct a stochastic process much like regular MCMC  except the normalizing function is adaptively approximated at each step of the algorithm by a stochastic approximation using the entire sample path up to that point. In order to make the normalizing function approximation approach  efficient, \cite{atch:lart:robe:2008} propose using an umbrella sampling approach with multiple particles at each iteration \citep{torr:vall:1977}. 

\subsubsection*{Pseudo-marginal Algorithms}

In the Metropolis-Hastings acceptance ratio, pseudo-marginal MCMC \citep{10.1214/07-AOS574} 
 replaces an intractable $\mathcal{L}(\theta\mid x)$ with its positive and unbiased Monte Carlo approximation $\hat{\mathcal{L}}(\theta\mid x)$. The resulting algorithm is useful for IL problems, and has the advantage of being asymptotically exact.

For INF problems, pseudo-marginal MCMC requires an unbiased approximation of $1/c(\theta)$. Although we can easily obtain an unbiased approximation for $c(\theta)$, $\hat{c}(\theta)$,  using importance sampling, obtaining an unbiased approximation of $1/c(\theta)$ is non-trivial. Note that $1/\hat{c}(\theta)$ is a consistent but biased approximation. To address this, \cite{10.1214/15-STS523} developed a geometric series correction method, called the Russian roulette algorithm. Under the pseudo-marginal framework, the algorithm is asymptotically exact, and assumptions are satisfied for general forms of  $h(x\mid\theta)$. However, implementing the stochastic truncation of the series requires multiple $\hat{c}(\theta)$s; considering that obtaining each $\hat{c}(\theta)$ requires Monte Carlo samples from $f(\cdot\mid\theta)$, the algorithm is computationally expensive. 

To speed up the algorithm, one might consider adapting surrogate likelihood approximations. For IL problems, \cite{drovandi2018accelerating} developed an approach to accelerate the pseudo-marginal algorithm by using a Gaussian process approximation of the log of an unbiased likelihood approximation. Developing asymptotically exact pseudo-marginal algorithms that are also computationally efficient is very challenging for INF problems. 

\subsubsection*{Bayesian Synthetic Likelihood}

Bayesian synthetic likelihood \citep{price2018bayesian} (BSL) is a likelihood-free algorithm that, like ABC, uses simulations from the probability model to modify the Metropolis-Hastings acceptance ratio. For each proposed $\theta^*$, the algorithm generates $m$ simulations $x_1^*,\dots, x_m^*$ from $f(\cdot \mid\theta^*)$, computes a summary statistic based on each simulation, $S(x_1^*),\dots, S(x_m^*)$, and then constructs a multivariate normal based on these statistics by simply using the sample mean and sample covariance; this is treated as an approximation to the distribution of the summary statistics. This multivariate normal is then used as a replacement for the true likelihood function in the Metropolis-Hastings acceptance ratio. The idea behind this algorithm is that if the summary statistics are approximately normal, this is a reasonable approximation. The algorithm is asymptotically inexact, like ABC, but it seems to outperform ABC in terms of computational efficiency in some examples \citep{price2018bayesian}, and so may be a useful addition to the toolkit for likelihood-free inference, and hence also intractable normalizing function problems.  

\subsubsection*{Variational Bayes}

Variational Bayes (VB) \citep{jordan1999introduction,bishop2006pattern}  approximates the posterior by minimizing the Kullback-Leibler divergence between $\pi(\theta\mid x)$ and a tractable distribution class, or equivalently maximizing the evidence lower bound. \cite{tran2017variational} develop VB for intractable likelihood problems by replacing the likelihood with unbiased importance sampling estimates. For INF problems, \cite{tan2020bayesian} develop two classes of VB methods for ERGMs with Gaussian posterior approximation. The first approach replaces the intractable likelihood with the adjusted pseudolikelihood \citep{bouranis2018bayesian} in optimizing the evidence lower bound. These adjustments correct the mode, curvature, and magnitude of the pseudolikelihood based on an affine transformation. The second one is a stochastic gradient ascent approach to optimize the evidence lower bound. Here, the gradient term is approximated through importance sampling estimates. 

VB approaches are attractive because they are potentially extremely fast, but both VB approaches of \cite{tan2020bayesian} need considerable tuning to be effective for a given distribution. 
The adjusted pseudolikelihood approach uses MLE and covariance estimates of sufficient statistics for an affine transformation, which requires Monte Carlo simulation from the model. Therefore, the quality of the preliminary iteration of MCMC-MLE \citep{geyer1992constrained,snijders2002markov} is crucial for the success of the algorithm. The stochastic gradient approach also uses the fitted results from the adjusted pseudolikelihood method as inputs for faster convergence; the performance of the algorithm can depend on them. Furthermore,  both VB approaches require simulations from the model, for adjusting pseudolikelihood and estimating the gradient respectively. However, if carefully tuned, VB approaches can be computationally faster than other approaches, including DMH, while providing reasonable posterior approximations. For example VB requires fewer auxiliary model simulations than DMH and the simulations can be parallelized. It is important to note that VB methods have so far only been developed for ERGMs \citep{tan2020bayesian}; it is unclear how to extend them to other models. 

\subsubsection*{Surrogate Likelihoods and Simulation-based Inference }

 A particularly exciting avenue for both INF and IL methods is exploring new machine learning methods that can be trained on samples to produce a surrogate model and likelihood function. The sampling can be run ahead of time, in parallel, and once the training is done an MCMC algorithm can use the surrogate likelihood. Hence, these algorithms can be very fast, though there are interesting challenges in terms of design---how to select the set of parameters for simulation---and architecture---how to construct the surrogate. For instance, \cite{sainsbury2023neural}  develop a neural network approximation to a Bayes estimator in the context of spatial data, and there are recent maximum likelihood approaches for expensive or intractable likelihoods \cite[cf.][]{walchessen2023neural,sainsbury2024likelihood}. There is a long history of using Gaussian processes in the complex computer models framework, that is, for inference where the model is a complex simulation model with no closed-form expressions \cite[cf.][]{gramacy2020surrogates,santner2003design}. Surrogates based on Gaussian processes have been developed for INF problems \cite[cf.][]{park2020function,drovandi2018accelerating}. For instance, in the LikeEm algorithm \citep{park2020function}, the normalizing function is approximated for a set of parameter settings using importance sampling techniques, then a Gaussian process is used to interpolate the approximated likelihood function over the entire parameter space. Once this pre-computing step is completed, the Gaussian process emulator is used instead of the likelihood function in the subsequent MCMC algorithm. This algorithm is fast but asymptotically approximate; it is studied as an example in Section \ref{sec:Applications}. The review in \cite{cranmer2020frontier} is targeted at IL problems rather than INF problems, and contains helpful descriptions of several  simulation-based surrogate likelihood methods. 
 We note that the area of simulation-based inference, which broadly includes any approach that uses machine learning methods on model simulations to do inference, is growing rapidly in terms of methods and applications; the methodology in this area is wide-ranging and has gone far beyond just directly approximating likelihoods. 

\section{Assessing Sample Quality}\label{sec:Assess}

Assessing the quality of samples from MCMC algorithms has been an active area of research for over three decades, with many theoretically justified and practical approaches (cf. \citet{roy2020convergence} and relevant chapters in this handbook). However, there are relatively few if any attempts at finding ways to assess the quality of samples produced by asymptotically inexact algorithms. Such diagnostics are important for ensuring the reliability of our results as well as for guidance for tuning our algorithm for any given problem. For example, the double Metropolis-Hastings algorithm requires determining the length of the inner and outer Markov chains; standard MCMC diagnostics are not useful for tuning this algorithm or assessing the quality of samples. Heuristics for tuning are often based on simulated examples and hence can be of limited  applicability to a particular problem. Determining which algorithm to prefer is also difficult without a good measure of sample quality, for example the exchange algorithm is  asymptotically exact but computationally  expensive while the double Metropolis-Hastings algorithm is faster but asymptotically inexact; which should we use for a given problem? 

There are two measures of sample quality provided by \cite{kang2024diagnostics}, the approximate curvature diagnostic (ACD) and the approximate inverse multiquadric kernel Stein discrepancy (AIKS). In the interest of brevity and because we find ACD to be more computationally expedient, we focus on ACD here. For more on AIKS and the kernel Stein discrepancy on which it is based, see \cite{kang2024diagnostics} 
 and \cite{gorham2017measuring} respectively. 

 \subsection{Curvature Diagnostic}
 The curvature diagnostic \citep{kang2024diagnostics} is inspired by maximum likelihood theory. The second Bartlett identity \citep{Bartlett1953a,Bartlett1953b} can be used to assess whether a model is correctly specified, that is, whether the data we observe are compatible with a particular probability model. The curvature diagnostic considers whether the samples drawn from an algorithm are compatible with the target posterior distribution $\pi(\theta \mid x)$. 
 Let $u(\theta)= \nabla_{\theta}\log \pi(\theta\mid x), H(\theta) = \frac{\partial}{\partial \theta} u(\theta)$, $J(\theta) = u(\theta) u(\theta)^\top$, and $d(\theta) = \textup{vech}[J(\theta) + H(\theta)]$, where $\textup{vech}(M)$ denotes the half-vectorization of the matrix $M$. 
If the samples from the algorithm are truly from the posterior, then $\textup{E}_{\pi}\{d(\theta)\} = 0$ by Bartlett's second identity. 
This leads to the following strategy: for samples $\{\theta^{(1)},\dots,\theta^{(n)}\}$ from the approximate algorithm, estimate the expectation as $ d_n = \frac{1}{n} \sum_{i=1}^n d(\theta^{(i)})$.
If the asymptotic distribution of the sample is the  target distribution $\pi(\theta \mid x)$, then $\sqrt{n} d_n \stackrel{d}{\to} \textup{N}(0, V)$ by the central limit theorem (CLT) for independent samples and MCMC CLT for samples from a Markov process. For independent samples, the unbiased and consistent approximation of the covariance matrix $V$ is calculated as $V_n = \frac{1}{n} \sum_{i=1}^n d(\theta^{(i)})d(\theta^{(i)})^\top$. For samples from a Markov chain, we can estimate $V$ using batch means estimator which is strongly consistent under some conditions \citep{Damerdji1994,Jones2006,Vats2019}. The curvature diagnostic is defined as $n d_n^\top V_n^{-1} d_n$.
This has an asymptotic $\chi^2(r)$ distribution, where $r = p(p+1)/2$ and $p$ is the dimension of $\theta$, if the asymptotic distribution of the sample is equal to the target posterior. The $1-\alpha$ quantile of the $\chi^2(r)$ can be used as a threshold for this diagnostic. A sample path for which the diagnostic value is below the threshold is considered to have an asymptotic distribution that is reasonably close to the target distribution.

\subsection{Approximate Curvature Diagnostic}
In the context of intractable normalizing function problems, $H(\theta)$ and $J(\theta)$ are intractable and hence need to to be approximated, leading to the approximate curvature diagnostic (ACD). \cite{kang2024diagnostics} describes how ACD is computed efficiently, providing theoretical justification for using ACD in place of CD.
The ACD is defined as $n \hat{d}_{n, N}^\top \hat{V}_{n,N}^{-1} \hat{d}_{n.,N}$ where $\hat{d}_{n, N}$ and $\hat{V}_{n,N}$ are the consistent estimates of $d_n$ and $V_n$, respectively. The approximations are obtained using auxiliary samples $\by^{(1)}, \dots, \by^{(N)}$ generated exactly from $f(\cdot \mid \theta)$ or generated by a Monte Carlo algorithm having $f(\cdot \mid \theta)$ as its stationary distribution. The self-normalized importance sampling \citep{tan2020bayesian} substantially speeds up the approximation step. Further details on the approximation are found in \citet{kang2024diagnostics}. In order to control standard errors and computational costs, it is important to construct these approximations carefully. The theoretical justifications may be summarized as follows: if the  asymptotic distribution of the sample is equal to the target distribution, then $\hat{d}_{n,N} \stackrel{\textup{a.s.}}{\to} d_n$ as $N \to \infty$ and $\hat{V}_{n,N} \stackrel{\textup{a.s.}}{\to} V$ as $n,N \to \infty$, so that the ACD converges to the curvature diagnostic. These results hold under reasonable conditions on the prior and likelihood; for instance they are satisfied for the challenging examples provided later in this chapter. The ACD has been shown to be effective in practice, as demonstrated on multiple challenging examples in \cite{kang2024diagnostics}. Further details on the application of ACD are in Section \ref{sec:Applications}. 

\section{Applications}\label{sec:Applications}
We showcase the application of five algorithms --- DMH \citep{Liang2010}, ALR \citep{atch:lart:robe:2008}, LikeEm \citep{park2020function}, VB \citep{tan2020bayesian},  and ABC-MCMC \citep{marjoram2003markov}. We study these algorithms in the context of three challenging examples, the Potts model, an exponential random graph model (ERGM), and an Ising network model, though not all the algorithms we consider are applied (or even applicable) to all the examples. The algorithms we have chosen are intentionally quite different from each other to provide readers with a sense of the various pros and cons of different classes of algorithms. 
The DMH is an auxiliary variable algorithm, while ALR is a function approximation approach that adaptively approximates the normalizing function at each step of the algorithm using the entire sample path up to that point. LikeEm is a function approximation approach that uses sampling to produce a Gaussian process surrogate to the likelihood. VB is a function approximation approach that uses optimization. ABC-MCMC is a subset of ABC algorithms that evaluates the Metropolis-Hastings acceptance ratio only if an auxiliary sample $x^{\ast}$ from the probability model given the proposed $\theta^{\ast}$ is close enough to the data $x$.
We find that all the algorithms we consider have some practical value in certain contexts, though not all the algorithms apply to all the examples we study. We find that our case studies provide insights about the algorithms and also demonstrate the value of the ACD measure of sample quality in tuning the algorithms, assessing whether the approximations are reasonable, and comparing the algorithms to each other. The source code can be downloaded from the following repository (\url{https://github.com/bokgyeong/HandbookMCMC}).

\subsection{Potts Model}\label{sec:potts}

\begin{figure}[!tbp]
    \centering
    \includegraphics[width = 0.45\textwidth]{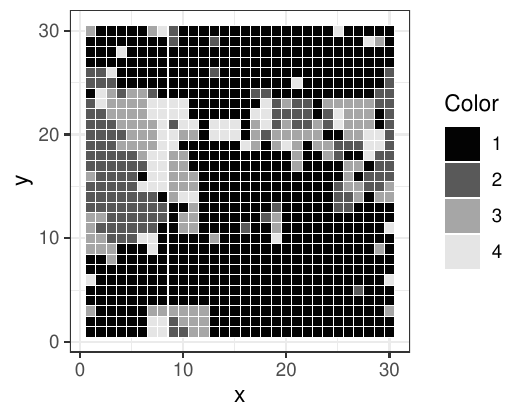}
    \caption{Multi-color image data simulated from the Potts model with $K$ = 4 and $\theta = \log(1 + \sqrt{4})$.}
    \label{fig:pottsData}
\end{figure}

The Potts model \citep{potts1952some}, a generalization of the Ising model \citep{lenz1920beitrag,ising1925beitrag}, provides an approach for modeling multi-colored images and hence discrete-valued spatial data on a lattice. For an $r \times s$ lattice $x$ with discrete values $x_{i} \in \{1, \dots, K\}$, the Potts model with   $\theta > 0$ has probability model
\begin{align*}
    f(x \mid \theta) &= \frac{1}{c(\theta)} \exp\left\{ \theta \sum_{i \sim j} I(x_i = x_j) \right\},
\end{align*}
where $i \sim j$ indicates neighboring elements, and $I(\cdot)$ denotes the indicator function. A larger value for $\theta$ produces a higher expected number of neighboring pairs that have the same color. Calculation of the normalizing function $c(\theta)$ requires summation over all $K^{rs}$ possible outcomes for the model, which is computationally infeasible even for lattices of moderate size. We simulated a 30 × 30 lattice with $K$ = 4 and $\theta = \log(1 + \sqrt{4})$ via 100,000 cycles of the Swendsen-Wang algorithm \citep{swendsen1987nonuniversal} using the R package \texttt{potts} \citep{potts}. The simulated dataset is presented in Figure~\ref{fig:pottsData}.

For this example, we consider the DMH, ABC-MCMC, and LikeEm algorithms that are described in Section~\ref{sec:Algs}. All algorithms are asymptotically inexact. We implement DMH with different numbers $m$ of Swendsen-Wang (inner sampler) updates. For ABC-MCMC, we consider different values $\epsilon$ for the threshold. The LikeEm algorithm has two tuning components: the number $d$ of particles and the sample size $N$ for importance sampling estimates. We fix $N$ = 500, which is a relatively large sample size, and study the performance of LikeEM with varying $d$. The particles are obtained by a short run of DMH with 90 cycles of inner updates. All algorithms were run for 300,000 iterations. We apply ACD to choose suitable values of inner sampler length $m$ for DMH, the threshold $\epsilon$ for ABC-MCMC, and particle size $d$ for LikeEm.
We used parallel computing to obtain 30 replications of ACD for each sample path. We assess the samples using the empirical mean of the replications. The threshold value of ACD is the 0.99 quantile of $\chi^2(1)$, which is 6.63.

\begin{figure}[!tbp]
    \centering
    \includegraphics[width = 0.9\textwidth]{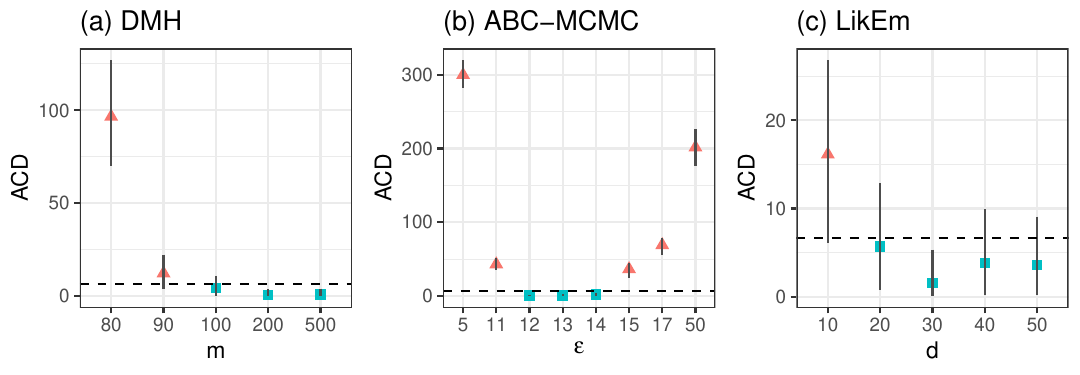}
    \caption{ACD results for the Potts model example. (a) ACD applied to samples generated from DMH with different numbers $m$ of (inner) Swendsen-Wang updates. (b) ACD applied to samples generated from ABC-MCMC with different values $\epsilon$ of the threshold. (c) ACD applied to samples generated from LikeEm with different numbers $d$ of particles. The dashed horizontal line represents the threshold value for the diagnostic. The triangle/square and vertical lines show the empirical mean and 95\% uncertainty interval, respectively, of 30 replications of the diagnostic. The red triangle and blue square indicate poor sample quality and good sample quality, respectively.}
    \label{fig:pottsACD}
\end{figure}

Figure~\ref{fig:pottsACD} shows the ACD values applied to each sample path. For DMH, ACD has its largest (worst) value at $m = 80$ and decreases (improves) as $m$ increases. ACD recommends $m$ of 100 or more for this example. ABC-MCMC has a trade-off in the choice of $\epsilon$; as it decreases, the approximation becomes more accurate at the expense of a higher rejection rate. ACD indicates that ABC-MCMC provides good sample quality for intermediate threshold values, i.e., $\epsilon$ = 12, 13, or 14. For LikeEm, ACD indicates that $d$ should be at least 20 to achieve a desirable approximation accuracy. LikEm with $d$ = 20 is the most computationally efficient of ACD-selected samples (i.e., DMH with $m \geq 100$, ABC-MCMC with $12 \leq \epsilon \leq 14$, and LikEm with $d \geq 20$), which took only 2.4 minutes.

\subsection{An Exponential Random Graph Model}\label{sec:ergm}

\begin{figure}[!tbp]
    \centering
    \includegraphics[width = 0.5\textwidth]{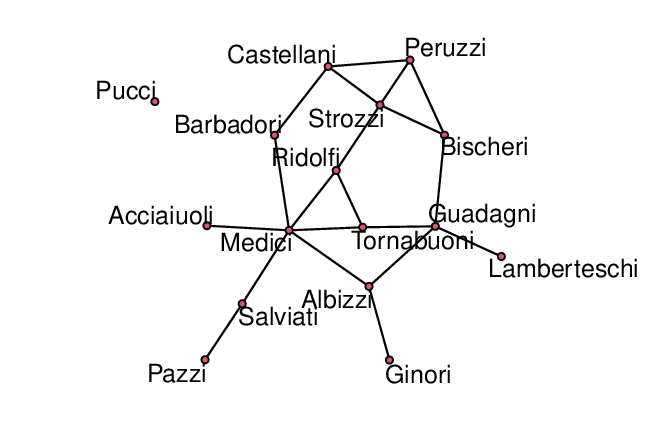}
    \caption{Florentine marriage network \citep{breiger1986cumulated} consists of 16 vertices and 20 undirected edges.}
    \label{fig:ergmData}
\end{figure}

Exponential random graph models (ERGM) \citep{robins2007introduction, hunter2008ergm} have been widely used to study the relationships between network nodes. Consider we have undirected network data $x$ with $n$ nodes. For all $i \neq j$, $x_{i,j}=1$ if the $i$th node and $j$th node are connected (neighbors), and $x_{i,j}=0$ otherwise. Then, the probability model of ERGM is
\begin{equation}
f(x\mid\theta) =  \frac{1}{c(\theta)}\exp\left\lbrace \theta_{1}S_{1}(x) + \theta_{2}S_{2}(x) \right\rbrace, \nonumber
\end{equation}
\begin{gather*}
S_{1}(x)=\sum_{i=1}^{n} \binom{x_{i+}}{1} ~~~~ S_{2}(x)=e^{0.2}\sum_{k=1}^{n-2}\left\lbrace 1-(1-e^{-0.2})^{k} \right\rbrace ESP_{k}(x)
\end{gather*}
where $S_{1}(x)$ and $S_{2}(x)$ indicate edges and the geometrically weighted edge-wise shared partnership (GWESP) statistics \citep{hunter2006inference}, respectively. The $ESP_{k}(\mathbf{x})$ term in the GWESP statistic indicates the number of connected $i,j$ pairs, where $i$ and $j$ have $k$ common neighbors. Therefore, GWESP can account for higher order transitivities with geometric weights. Evaluation of the normalizing function $c(\theta)$ is intractable because it requires summation over all $2^{n(n-1)}$ possible configurations in the network. We study the Florentine marriage dataset \citep{breiger1986cumulated}, which describes the marriage alliance networks among 16 Florentine families and is shown in Figure~\ref{fig:ergmData}.


\begin{figure}[!tbp]
    \centering
    \includegraphics[width = 0.9\textwidth]{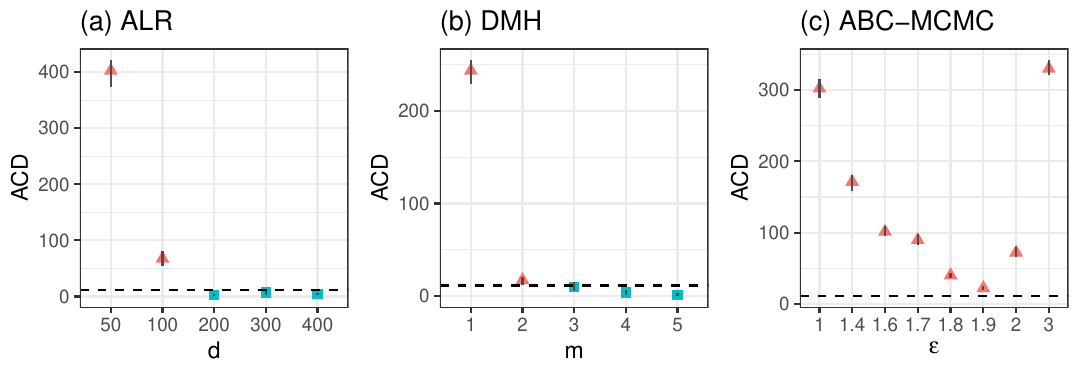}
    \caption{ACD results for the ERGM example. (a) ACD applied to samples generated from ALR with different numbers $d$ of particles. (b) ACD applied to samples generated from DMH with different numbers $m$ of (inner) Gibbs updates. (c) ACD applied to samples generated from ABC-MCMC with different values $\epsilon$ of threshold. The triangle/square and vertical line show the empirical mean and 95\% uncertainty interval, respectively, of 30 replications of the diagnostic. The red triangle and blue square indicate poor sample quality and good sample quality, respectively.}
    \label{fig:ergmACD}
\end{figure}

For this example we consider ALR, DMH, ABC-MCMC and VB algorithms that are described in Section~\ref{sec:Algs}. Except for ALR, all algorithms are asymptotically inexact. The performance of ALR relies heavily on the set of particles which should cover the important region of the parameter space to provide good posterior samples. We implement ALR with different numbers $d$ of particles.
For DMH and ABC-MCMC, we consider different numbers $m$ of Gibbs (inner) updates and different values $\epsilon$ of threshold, respectively. The algorithms were run for 200,000 iterations. We tune the VB algorithm following \citet{tan2020bayesian} and obtain an approximate posterior distribution. We generate 100,000 samples independently from the approximate posterior. We apply ACD to each sample path for assessing sample quality. We used parallel computing to obtain 30 replications of ACD for each sample path. We assess the samples using the empirical mean of the replications. The threshold value of ACD is the 0.99 quantile of $\chi^2(3)$, which is 11.34.

\begin{figure}[!tbp]
    \centering
    \includegraphics[width = \textwidth]{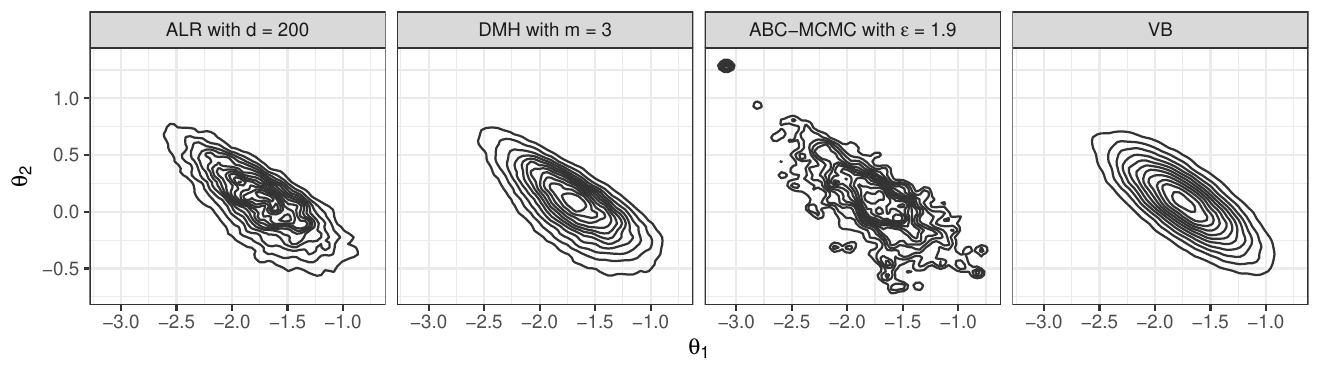}
    \caption{The estimated joint posterior density of $\theta_1$ and $\theta_2$ for each algorithm for the ERGM example.}
    \label{fig:ergmDen}
\end{figure}

Figure~\ref{fig:ergmACD} shows the ACD values for each posterior sample path. For ALR and DMH, ACD generally decreases (improves) as $d$ and $m$ increase, respectively. ACD recommends $d$ of at least 200 for ALR and $m = 3$ or more for DMH. ACD indicates that ABC-MCMC produces poor sample quality for all threshold values examined in this example. ACD for VB is 956.31 which is much greater than the ACD threshold value of 11.34. This implies that VB does not perform well even though we carefully tune the algorithm. 

Figure~\ref{fig:ergmDen} shows the estimated joint posterior density for each algorithm with a tuning parameter value chosen by ACD. We observe that ALR and DMH provide similar joint posterior densities. ABC-MCMC seems to mix poorly. VB fails to correctly capture relationship between two parameters. DMH with $m$ = 3 is most computationally efficient of ACD-selected samples (i.e., ALR with $d \geq 200$ and DMH with $m \geq 3$), which took only 1.6 minutes.


\subsection{An Ising Network Model: Applications to Verbal Aggression Data}\label{sec:ising}

Ising network models \citep{van2014new} are undirected graphical models that can describe interactions among binary responses. Consider binary item response data $x$ with $n$ responses to $p$ items. For all $i,j$, $x_{i,j}=1$ if the $i$th respondent answers the $j$th item correctly (or positively), and $x_{i,j}=0$ otherwise. The Ising network model with parameters $\theta=(\beta,\gamma)$ has probability model
\begin{equation}
f(x\mid\theta) = \frac{1}{c(\theta)}\exp\left\{ \sum_{j=1}^{p}\beta_{j}\sum_{i=1}^{n}x_{ij} + \sum_{j<k}\gamma_{jk}\sum_{i=1}^{n}x_{ij}x_{ik}  \right\},
\label{isingnet}
\end{equation} 
where $\beta_j$ is an item easiness parameter, and $\gamma_{jk}$ is a pairwise interaction parameter between item $j$ and $k$. Calculation of the normalizing function $c(\theta)$ requires summation over all $2^{np}$ possible configurations, which is intractable. Furthermore, the model includes $p + \binom{p}{2}$ parameters, which are high-dimensional. Here we analyze the item responses to a questionnaire on verbal aggression \citep{de2004explanatory}. All items are about verbally aggressive reactions in a frustrating situation, and we focus on studying 12 {\it{want}} behavior mode items as follows:
\begin{enumerate}[(1)]
\item A bus fails to stop for me. I would want to curse. 
\item A bus fails to stop for me. I would want to scold. 
\item A bus fails to stop for me. I would want to shout. 
\item I miss a train because a clerk gave me faulty information. I would want to curse. 
\item I miss a train because a clerk gave me faulty information. I would want to scold. 
\item I miss a train because a clerk gave me faulty information. I would want to shout.
\item The grocery store closes just as I am about to enter. I would want to curse.  
\item The grocery store closes just as I am about to enter. I would want to scold.    
\item The grocery store closes just as I am about to enter. I would want to shout.          
\item The operator disconnects me when I had used up my last 10 cents for a call. I would want to curse.  
\item The operator disconnects me when I had used up my last 10 cents for a call. I would want to scold.    
\item The operator disconnects me when I had used up my last 10 cents for a call. I would want to shout.          
\end{enumerate}
All responses were dichotomized to have binary values (either $1=$“yes” or $0=$“no”). The data include $n=316$ respondents for the $p=12$ items described above; the resulting Ising network model has  $12 + \binom{12}{2} = 78$ parameters.

In order to detect significant interactions among items, we apply the spike and slab DMH sampler \citep{park2022ising} that poses spike and slab priors for $\theta$ as follows: 
\begin{equation}
\begin{split}
\theta_i \mid \lambda_i, \sigma^2, \omega &\overset{\text{ind}}{\sim} \lambda_i N(0, \omega^2\sigma^2) + (1-\lambda_i) N(0, \sigma^2),\\
\lambda_i  &\overset{\text{iid}}{\sim} \mbox{Bernoulli}(1/2),\\
1/\sigma^2 &\sim \mbox{Uniform}(4,100),\\ 
\omega   &\sim 1 + Y,\\
Y &\sim \mbox{Gamma}(1, 1/100),
\end{split} 
\label{spike}
\end{equation}
where $\lambda_i$ is a latent variable indicating whether $\theta_i$ is included in the model ($\lambda_i=1$) or not ($\lambda_i = 0$), and $\sigma^2$ and $\omega$ control the variances of spike and slab distributions, respectively. Inference for the parameters is carried out by the DMH algorithm. 
DMH is a practical option for such high-dimensional hierarchical models with intractable normalizing functions. DMH can be implemented in general cases once we have an inner sampler. Other algorithms like VB and LikeEm can be hard to tune or impractical due to the high dimensional parameter space. 
For DMH, the length $m$ of the inner sampler should be carefully chosen. 
However, ACD is impractical as it requires approximating the 3,081 $\times$ 3,081 covariance matrix of $d(\theta)$. To provide a reliable approximation to the covariance matrix, we need long posterior sample paths, resulting in huge demands on memory.  \cite{park2022ising} study the performance of the spike and slab DMH sampler with different lengths $m$ of inner sampler. Following the suggestion in \cite{park2022ising}, we run the algorithm with $m = 10n$, where $n$ is the number of respondents. 

\begin{figure}[tbp]
\begin{center}
\includegraphics[width=0.45\textwidth]{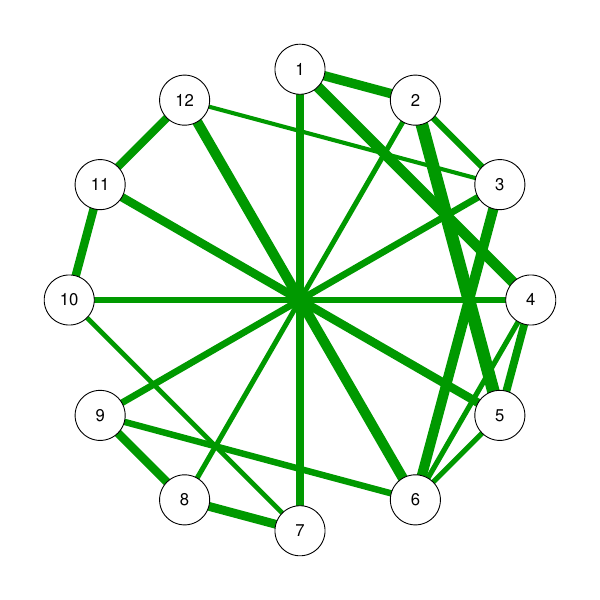}
\end{center}
\caption[]{An estimated network structure for the verbal aggression data. The width of the lines indicates the strength of the connection between the relevant items; thicker lines indicate stronger interaction between items.}
\label{IsingNet}
\end{figure}

Of the 66 pairwise interaction parameters $\gamma_{jk}$, the spike and slab DMH shrinks 45 interaction parameters toward 0 and provides positive values for the others. Figure~\ref{IsingNet} illustrates the resulting estimated network structure. 
We observe that $\gamma_{2,5}$, $\gamma_{1,4}$, and $\gamma_{3,6}$ have the greatest posterior means, indicating the strongest positive interactions. The strongest positive interaction is observed between items (2) (“A bus fails to stop for me. I would want to scold”) and (5) (“I miss a train because a clerk gave me faulty information. I would want to scold”), which makes sense because both items are about scold behavior. The second strongest positive interaction appears between items (1) (“A bus fails to stop for me. I would want to curse”) and (4) (“I miss a train because a clerk gave me faulty information. I would want to curse”); both items are about curse behavior. Lastly, the third strongest positive interaction occurs between items (3) (“A bus fails to stop for me. I would want to shout”) and (6) (“I miss a train because a clerk gave me faulty information. I would want to shout”), indicating that shout behaviors are also strongly connected. This example illustrates the shortcomings of the state-of-the-art in computing for INF problems. Although DMH appears to be practical in this context, the closest we have to assurance about the quality of sample-based inference is based on a heuristic that relies on experience. 



\section{Summary}\label{sec:Summary}

Inference in the presence of intractable normalizing functions is an exciting computing problem with lots of room for creativity. While it is difficult to provide an exhaustive review, we hope we have provided a reasonably broad perspective on key ideas for this problem. Because of the number of algorithms and the vast difference among them, as well as the scarcity of theory and heuristics for comparing them, it can be difficult to determine which algorithm to apply for any given situation.  
The recent measures of sample quality in \cite{kang2024diagnostics} may be helpful in this regard, as they have some potential for evaluating not only the quality of samples from a particular algorithm but for comparing algorithms from very different categories, including asymptotically exact and inexact algorithms. We find through our study, using the ACD diagnostic \citep{kang2024diagnostics} to measure sample quality where possible, that the double Metropolis-Hastings algorithm is quite effective and broadly applicable when used in tandem with ACD, and also has the advantage of being easier to code than most algorithms for Bayesian inference with intractable normalizing functions. Of course we must add the caveat that our study of algorithms is necessarily limited to a few challenging examples and a small set of interesting algorithms and hence we do not claim that our results will necessarily hold across the enormous spectrum of intractable normalizing function problems. 

\subsection*{Potential Directions for Research}

This chapter has provided a sampling of the many creative ideas that have emerged in recent years for inference with intractable normalizing functions and intractable likelihoods. There are clearly many open practical and theoretical issues. The computational complexity of algorithms for intractable likelihood  and intractable normalizing function problems makes them computational expensive, and sometimes impractical, for many real applications. Examples of such problems include 
interaction point process models \citep[cf.][]{goldstein2015attraction}, network psychometrics \citep[cf.][]{van2014new}, exponential random graph models \citep[see][]{Robins2007,Hunter2006}, and mixed graphical models \citep[cf.][]{Lauritzen1989,Lee2015,Cheng2017}
, and mean-parameterized Conway--Maxwell--Poisson models \citep[cf.][]{huang2017,Huang2019,kang2024comp}.
 A major computational bottleneck for most algorithms is the need to generate a large number of expensive auxiliary samples from the probability model. Variational methods offer a promising alternative because the sampling can be parallelized and done in advance; they are potentially very fast but the amount of problem-specific tuning they require to make them work well can make them impractical for many settings; finding ways to make VB work well in these scenarios is an open area for research. While they have been around for a while, composite likelihood methods avoid the need for auxiliary simulation, can be relatively easy to implement, and seem to work well in certain contexts. It may be worth exploring composite likelihood approaches for a variety of INF problems though of course composite likelihood methods are of limited applicability, mostly confined to applications in spatial and network models. Finally, more generalized likelihood approximation approaches show much promise. We show in a real example in Section \ref{sec:Applications} that a Gaussian process-based likelihood approximation can be much more efficient while remaining just as reliable as other algorithms for inference. 
 Approximations based on machine learning techniques have advanced a great deal beyond Gaussian processes \citep[cf.][which describes a variety of approximation methods]{cranmer2020frontier}.
 Hence, the application of fast machine learning techniques to likelihood function approximations is a promising avenue for future research as well, with many interesting challenges to address in terms of theory and applications. 

An important problem is assessing the quality of approximations obtained from asymptotically inexact algorithms. The ACD and AIKS methods \citep{kang2024diagnostics} provide a nice way to measure sample quality for asymptotically inexact algorithms, but do not actually provide a way to measure the error in the approximation of particular quantities of interest, typically taking  the form of a specific expectation $E_{\pi}g(X)$ for a given target $\pi$ and real-valued function $g$. As is well known to MCMC users, the requisite length of the chain varies depending on the required accuracy (MCMC standard error) and the specific targeted function $g$ \citep{fleg:hara:jone:2008}.  For instance approximating tail probabilities, $g(x)=I(x>c)$ for large $c$ and higher moments, $g(x) = x^2$, tend to require more samples than simple expectations ($g(x)=x$). There are, to our knowledge, no methods  for measuring the accuracy of approximations for these different quantities for an asymptotically inexact algorithm when the normalizing function or likelihood function is intractable. The diagnostics we recommend can also be computationally demanding, as is apparent from the challenge of applying ACD in the Ising network models example; developing computationally cheaper alternatives will be of value. Another major open problem is measuring the quality of samples produced by  inexact algorithms when the entire likelihood function is  intractable, for example in ABC and BSL algorithms applied to IL problems, though we found some ABC-specific tools that may be useful  \citep{prangle2014diagnostic,rendsburg2022discovering}. 

\section*{Acknowledgments}
JP was partially supported by the National Research Foundation of Korea (2020R1C1C1A0100386814, RS-2023-00217705) and the ICAN (ICT Challenge and Advanced Network of HRD) support program (RS-2023-00259934) supervised by the IITP (Institute for Information \& Communications Technology Planning \& Evaluation).

\bibliographystyle{apalike} 
\bibliography{short, references, refsjohn}

\end{document}

%% file: preamble.tex
\newcommand\by{{\bf y}}

